\documentclass[journal]{IEEEtran}

\usepackage{graphicx}
\usepackage{float}
\usepackage{dirtytalk}
\usepackage{footmisc}
\usepackage{amsmath}
\usepackage{hyperref}

\ifCLASSINFOpdf
  
\else
  
\fi

\begin{document}

\title{Similarity-Based Predictive Maintenance Framework for Rotating Machinery}

\author{Sulaiman~Aburakhia,~\IEEEmembership{Student Member,~IEEE,}
       Tareq~Tayeh,
        Ryan~Myers, and~Abdallah~Shami,~\IEEEmembership{Senior~Memeber,~IEEE}%

\thanks{S. Aburakhia, T. Tayeh, and A. Shami are with the Department of
Electrical and Computer Engineering, Western University, London, ON
N6A 3K7, Canada (e-mail: saburakh@uwo.ca; ttayeh@uwo.ca; abdallah.shami@uwo.ca). R. Myers is with the National Research Council Canada, London, ON N6G 4X8, Canada (e-mail: ryan.myers@nrc-cnrc.gc.ca)}}

\markboth{}%
{Shell \MakeLowercase{\textit{et al.}}: Bare Demo of IEEEtran.cls for IEEE Journals}

\IEEEoverridecommandlockouts
\IEEEpubid{\makebox[\columnwidth]{978-1-6654-8237-0/22\$31.00~\copyright2022 IEEE \hfill} \hspace{\columnsep}\makebox[\columnwidth]{ }}

\maketitle

\IEEEpubidadjcol

\begin{abstract}
Within smart manufacturing, data driven  techniques are commonly adopted for condition monitoring and fault diagnosis of rotating machinery. Classical approaches use supervised learning where a classifier is trained on labeled data to predict or classify different  operational states of the machine. However, in most industrial applications, labeled data is limited in terms of its size and type. Hence, it cannot serve the training purpose. In this paper, this problem is tackled by addressing the classification task as a similarity measure to a reference sample rather than a supervised classification task. Similarity-based approaches require a limited amount of labeled data and hence, meet the requirements of real-world industrial applications. Accordingly, the paper introduces a similarity-based framework for predictive maintenance (PdM) of rotating machinery. For each operational state of the machine, a reference vibration signal is generated and labeled according to the machine's operational condition. Consequentially, statistical time analysis, fast Fourier transform (FFT), and short-time Fourier transform (STFT) are used to extract features from the captured vibration signals. For each feature type, three similarity metrics, namely structural similarity measure (SSM), cosine similarity, and Euclidean distance are used to measure the similarity between test signals and reference signals in the feature space. Hence, nine settings in terms of feature type-similarity measure combinations are evaluated. Experimental results confirm the effectiveness of similarity-based approaches in achieving very high accuracy with moderate computational requirements compared to machine learning (ML)-based methods. Further, the results indicate that using FFT features with cosine similarity would lead to better performance compared to the other settings.

\end{abstract}


\IEEEpeerreviewmaketitle

\section{Introduction}

\IEEEPARstart{P}{redictive} maintenance (PdM) approaches have been widely adopted in recent years for maintenance management in rotating machinery. PdM relies on continuously monitoring the equipment's condition, and actions for maintenance are predicted based on the equipment's actual condition. PdM involves two main tasks: First,  extracting useful features from equipment-related data---such as vibration signals generated by rolling bearings of the rotating machinery--- that can describe the process integrity well with high sensitivity to any changes within the process. The second task involves utilizing the extracted features to classify or predict normal and abnormal operational conditions with high accuracy. Common feature extraction methods include time domain analysis, frequency domain analysis, and time-frequency domain analysis \cite{Attoui1}--\cite{Nayana}. In time-domain analysis, common statistical properties of the signal such as kurtosis, skewness, crest factor, peak...etc. are used as features. Although, time domain analysis is considered a simple approach to extract features, it has low sensitivity to process variations.  Frequency domain analysis such as FFT allows the extraction of spectral-related features that are sensitive to the variations in operational conditions. However, frequency domain analysis has no resolution in the time domain. Moreover, its application is limited to stationary signals. On the other hand, time-frequency domain analysis has better temporal and frequency localization compared with the Fourier analysis. Common time-frequency domain analysis methods include, STFT, wavelet transform, and Hilbert–Huang transform (HHT). Regarding the classification task, classical methods utilize supervised learning techniques to train a classifier on the extracted features. These methods usually require large-sized labeled data to fulfill training requirements. However, in most real-world situations, the available labeled data is limited in its size. Moreover, it is difficult to have labeled data that can model all possible classes or operational conditions sufficiently. For example, it is possible to obtain sufficient samples that can model standard or normal conditions. While on the other hand, samples of abnormal conditions are usually not abundant and insufficient to model all possible abnormal operational conditions. Similarity-based approaches \cite{Attoui1}\cite{Attoui2}\cite{parun}--\cite{tar} offer an alternative solution to perform classification tasks with limited labeled data. In contrast to supervised learning, similarity-based techniques achieve classification tasks by measuring the similarity between a given test sample and a labeled reference sample, which can be achieved using very limited labeled data. This paper introduces a similarity based PdM framework for rotating machinery. The monitoring of process integrity is achieved by continuously analysing the vibration signal generated by rolling-element bearing. For each operational condition, a reference signal is generated and labeled according to the current operational condition. Then, signal processing-based methods are used to extract features from captured vibration signals. Accordingly, similarity between test signals and reference signals is measured in the feature space to predict different operational conditions of test signals. The main contributions of the paper are:
\begin{itemize}
     \item Introducing a similarity-based framework for condition monitoring of rotating machinery. The main aspects of the framework are feature extraction and similarity-based classification.
    \item Three types of features, namely, time, frequency and time-frequency features are extracted from vibration signals.
    \item For each feature type, three similarity metrics are used for similarity-based classification. The three metrics are SSM, cosine similarity, and Euclidean distance.
\end{itemize}

The paper is outlined as follows: The next section provides a review on related work. Methodology and framework are introduced in Section 3 and section 4, respectively.  Section 5 presents the dataset and the experimental setup for performance evaluation while Section 6 discusses the results. The paper is finally concluded in Section 7.

\section{Related Work}
Data-driven techniques are commonly adapted to perform PdM of rotating machinery using the vibration signals of rolling bearings \cite{saburakhia}. However, majority of the proposed techniques in the literature rely on ML-based classification and less attention is paid to similarity-based approaches. In \cite{Nayana}, a comparative study to evaluate the effectiveness of statistical time domain features with several classifiers is attempted. Results show that the extracted features are effective in identifying bearing faults. Further, it was found that the accuracy of classification increases as the length of captured signal increases. However, increasing the length of captured signals would increase computational requirements. Moreover, it could delay triggering the faults. In \cite{Attoui1}, \cite{Attoui2}, and \cite{parun}, frequency domain features and time-frequency domain features along with similarity-based classification approaches are utilized for rolling bearing condition monitoring. In \cite{Attoui1}, a labeled reference signal from each operation condition is generated in first place. Consequentially, FFT is used to extract the features from the signals, and different operational conditions are determined by applying a proposed statistical similarity measure between test samples and reference samples. In \cite{Attoui2}, a similar approach is used. However, in contrast to \cite{Attoui1}, SSM \cite{Bovik} is adapted to measure the  similarity. Further, wavelet packet decomposition (WPD) is used to improve system robustness against noise and in the same time, increase its sensitivity to local differences in vibration signals. Consequentially, STFT is used to extract features from reconstructed signals. In \cite{parun}, the spectrogram images of the test vibration signals are compared with spectrogram images of normal baseline vibration signals using SSM. Consequentially, normal and faulty vibration signals are classified by setting a threshold on the resulting SSM scores. The main aspect in \cite{Attoui1}, \cite{Attoui2}, and \cite{parun} is the use of signal processing for features extraction along with similarity-based classification, which eliminates the need for machine learning-based trained classifiers. In this paper, we introduce a framework for similarity-based condition monitoring of rotating machinery, and use statistical time properties, FFT, and STFT to extract the features from variation signals. Further, in contrast to \cite{Attoui1}, \cite{Attoui2}, and \cite{parun} where statistical and structural similarities are used, we also use cosine similarity and Euclidean distance---which are less complex compared to SSM---to measure the similarity between reference and test samples.

\begin{figure}[t]
\centerline{\includegraphics[width=0.5\textwidth]{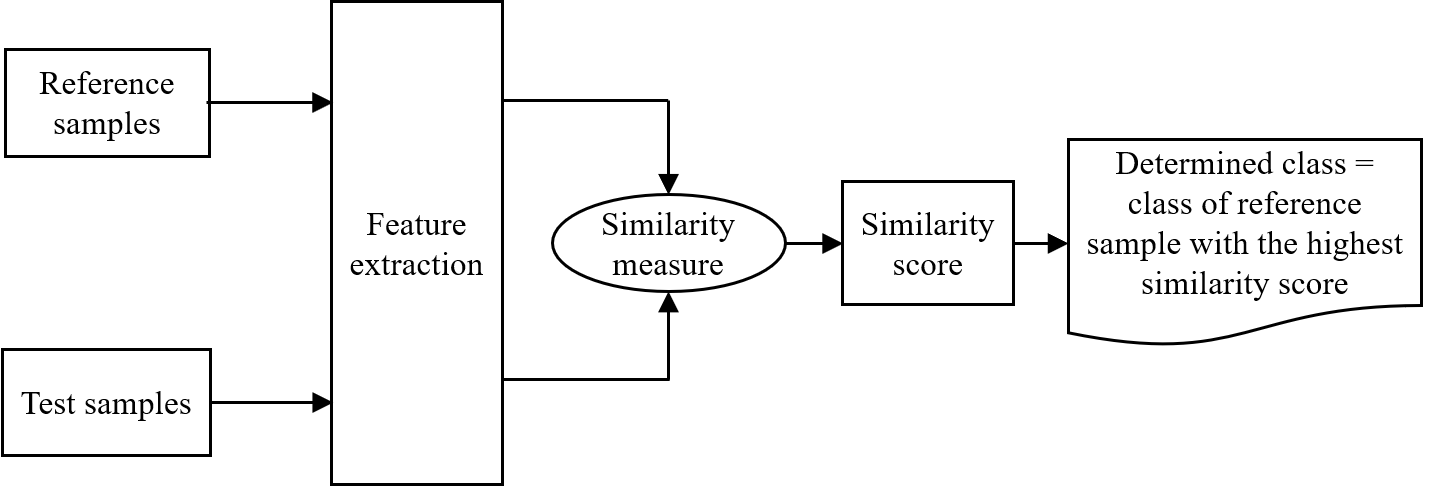}}
\caption{Similarity-based classification framework.}
\label{fig1}
\end{figure}

\section{Methodology}
Fig.\ref{fig1} shows a flowchart of the similarity-based classification framework. In the first stage, discriminative features are extracted from labeled reference samples and test samples. Consequentially, similarity between reference samples and test samples is measured in the feature space. Finally, the classification of different operational conditions is achieved by evaluating the resulting similarity scores. In contrast to machine learning-based approaches, the framework utilizes similarity scores to determine operational conditions instead of a trained classifier. The main aspects of the framework are feature extraction and similarity measure. Extracted features should be selected so that they satisfy two main conditions:
\begin{itemize}
\item Describe the inherent characteristics of all operational conditions \say{classes} in the data.
\item Have high-discrimination degree between the different operational conditions in the data.
\end{itemize}
Once features are extracted, the similarity measure is applied to quantify the similarity between reference samples and test samples in the feature space. To perform similarity measure, a reference sample from each operational condition (class) should be available. Since the similarity measure provides a quantitative value, it can be used to assess the probability that the reference sample and test sample belong to the same operational condition. The higher the similarity, the higher the probability that they belong to the same condition. This can be expressed mathematically as follows:
For a given test sample $X_{n}, n = 1,..,N$ and a reference sample $Y_{i}, i =1,..,m$, the similarity score $s_{ni}$ can be defined as:
\begin{equation}
s_{n,i}=S(F(X_{n}), F(Y_{i}))
\end{equation}
where $N$ is number of test samples, $m$ is number of operational conditions or classes, $i$ is the class of the reference sample, $S$ is the applied similarity measure, and $F$ denotes the feature extraction function.
The operational condition $C_{X_n}$ of $X_n$ can be determined according to the below equation;
\begin{equation}
C_{X_n}= i \textbf{   if   } s_{n,i}= M(S)
\end{equation}
where,
\begin{equation}
M(S)= M\{s_{n,1}, s_{n,2},...,s_{n,m}\},
\end{equation}
$M\{ \}$ denotes the "maximum" or "minimum" operation depending on the similarity measure type.

\section{Predictive Maintenance Framework for Condition Monitoring of Rotating Machinery}
In this section, the similarity-based framework is presented in details. The framework preforms PdM by continuously analysing the vibration signal of rolling-element bearing. To simulate a noisy environment, vibration signals are corrupted with additive white Gaussian noise (AWGN) at different signal-to-noise ratio (SNR) levels. To increase the system's robustness against the noise, the signals are denoised using WPD. For each operational condition, a labeled reference signal is generated. Its label represents the class to which the signal belongs in terms of operational condition. To simulate the effects of a noisy environment, reference and test signals are corrupted by AWGN at desired SNR. In the next step, signal denoising is applied. Consequently, features are extracted from denoised signals. In the final stage, similarity in the feature space is measured between each test signal and all labeled reference signals. Accordingly, for a given test signal, its  determined class will be the class of the reference signal with the highest similarity score. 

\subsection{Signal Denoising}
Generally, noise  presence in the signal is characterized as high frequency components. Thus, signal denoising can be accomplished by decomposing the signal using WPD technique and filtering-out detail coefficients associated with higher frequency sub-bands. This can be achieved by thresholding the detail coefficients so that coefficients below the threshold are set to zero. The denoised signal is then reconstructed using the approximation and the thresholded detail coefficients. In this paper, Daubechies 4 (db4) wavelet is used to decompose the noisy signals and the soft thresholding function \cite{Jing} is applied to the details coefficients. The threshold value is determined according to the below formulas \cite{ddonoho}:
\begin{equation}
threshold=\sigma\sqrt{2\log (N)/N},
\end{equation}
\begin{equation}
\sigma = \frac{\textit{median}(|w_k|)}{0.6745},
\end{equation}
where $N$ is signal length, and $|w_k|$ are wavelet coefficients.

\subsection{Feature Extraction}
\begin{figure*}[t]
\centerline{\includegraphics[width=0.9\textwidth]{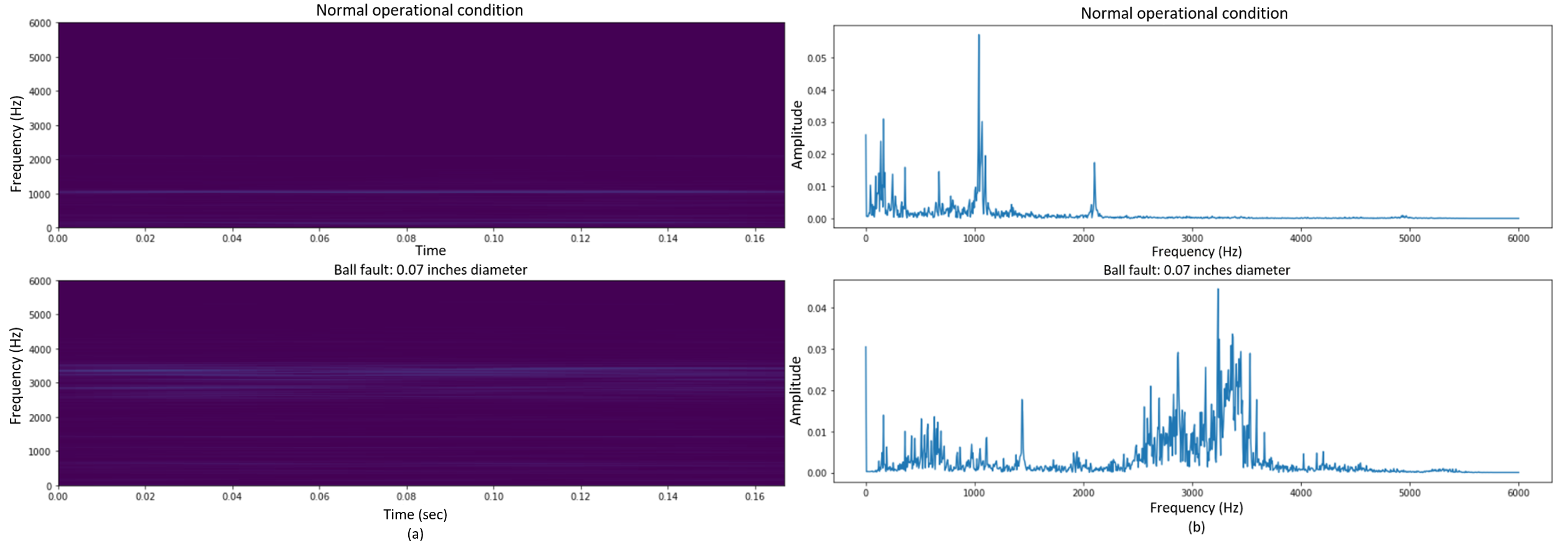}}
\caption{(a) STFT and (b) FFT of normal (healthy) operational condition and a faulty operational condition (ball fault of 0.07 inches in diameter).}
\centering
\label{fig3}
\end{figure*}

Time domain analysis, frequency domain analysis, and time-frequency domain analysis are used to extract features from vibrations signals. The aim here is to evaluate and compare the performance of time, frequency , and time-frequency features in a noisy environment. The statistical proprieties of vibration signals will be used as time domain features. FFT will be used to obtain spectral components of the signals and the positive part of the spectrum will be used as frequency domain features. For time-frequency domain analysis, STFT will be used. Fig \ref{fig3} shows FFT and STFT contents of bearing's vibration signal of a normal operational condition along with a faulty operational condition. As shown,  frequency components of the vibration signal are very sensitive to changes in operational conditions. Thus, in contrast to time-domain features, STFT and FFT  provide very useful features with high discrimination degree for the classification of different operational conditions. 

\subsection{Similarity Measure}
To evaluate the similarity between test signals and reference signals in the feature space, three different similarity measures are used and compared in terms of classification accuracy. The three measures are: cosine similarly, Euclidean distance, and SSM \cite{Bovik}. The cosine similarity $C_s$ between two vectors X and Y of length $N$ is;
\begin{equation}
C_s(X,Y)=\frac{\sum_{n=1}^{N}X_n\cdot Y_n}{\sqrt{\sum_{n=1}^{N} X_n^{2}}\cdot \sqrt{\sum_{n=1}^{N} Y_n^{2}}}
\end{equation}
For a given test signal $X_n$ and $m$ reference signals $Y_{i}, i =1,..m$, the class of $X_n$ is determined as follows:
\begin{equation}
\textbf{ Class of } X_n = i \textbf{   if   } C_s(F(X_n), F(Y_i)) = min\{D\}
\end{equation}
where,
\begin{equation}
\begin{split}
D & = \{C_s(F(X_n),F(Y_1)),\\
 & C_s(F(X_n),F(Y_2)),..,C_s(F(X_n),F(Y_m))\}.
\end{split}
\end{equation}\

The Euclidean distance between two vectors X and Y of length $N$ is:
\begin{equation}
{d \left({\it X },{\it Y }\right)=\sqrt{\left({\it X_1}{\it Y_1 }\right)^{2}+...+\left({\it X_N }{\it Y_N }\right)^{2}},\mathop{\rm  }n \mathop{\rm  }=1,..N ,\mathop{\rm  }\mathop{\rm}}
\end{equation}

For a given test signal $X_n$ and $m$ reference signals $Y_{i}, i =1,..m$, the class of $X_n$ is determined as follows:
\begin{equation}
\textbf{ Class of } X_n = i \textbf{   if   } d(F(X_n), F(Y_i)) = min\{D\}
\end{equation}
where,
\begin{equation}
\begin{split}
D & = \{d(F(X_n),F(Y_1)),\\
 & d(F(X_n),F(Y_2)),..,d(F(X_n),F(Y_m))\}.
\end{split}
\end{equation}\

The SSM algorithm \cite{Bovik} is an image quality assessment metric; it provides a perceptual metric to quantify the degradation in image quality caused by image processing such as compression, transmission..etc. The algorithm requires two input images, a processed image and its reference image. The structural similarity between the input images is then evaluated by comparing luminance, contrast, and structure of the two input images.The output is a value between 0 and 1 that quantifies the quality of the processed image with respect to the reference image. The higher the value, the higher the quality.\ 
Accordingly, SSM is defined as \cite{Bovik}:
\begin{equation}
SSM(X,Y) = \frac{(2\mu_{x}\mu_{y}+C_{1})(2\sigma_{xy}+C_{2})}{(\mu_{x}^2+\mu_{y}^2+C_{1})(\sigma_{x}^2+\sigma_{y}^2+C_{2})}
\end{equation}
where, $\sigma_{{\it xy}}, \mu_{x},\mu_{y}, \sigma_{x},\sigma_{y}$ are covariance, means and standard deviations of vectors $X$ and $Y$; calculated over a window of size $w$, in this paper, the window size is set to 7. $C_{1,2}$ are arbitrary constants to avoid unstable output when either $(\mu_{x}^2+\mu_{y}^2+)$ or $(\sigma_{x}^2+\sigma_{y}^2)$ is very close to zero.
For a given test signal $X_n$ and $m$ reference signals $Y_{i}, i =1,..m$, the class of $X_n$ is determined as follows:
\begin{equation}
\textbf{ Class of } X_n = i \textbf{   if   } SSM(F(X_n), F(Y_i)) = max\{D\}
\end{equation}
where,
\begin{equation}
\begin{split}
D & = \{SSM(F(X_n),F(Y_1)),\\
 & SSM(F(X_n),F(Y_2)),..,SSM(F(X_n),F(Y_m))\}.
\end{split}
\end{equation}

In this paper, SSM is used to measure the similarity between reference samples and test samples in the feature space where $X$ and $Y$ represent the extracted features of test samples and reference samples, respectively.

\begin{table}
  \caption{Bearing test data.}
  \label{dataset_table}
  \centering
  \includegraphics[scale = 0.4]{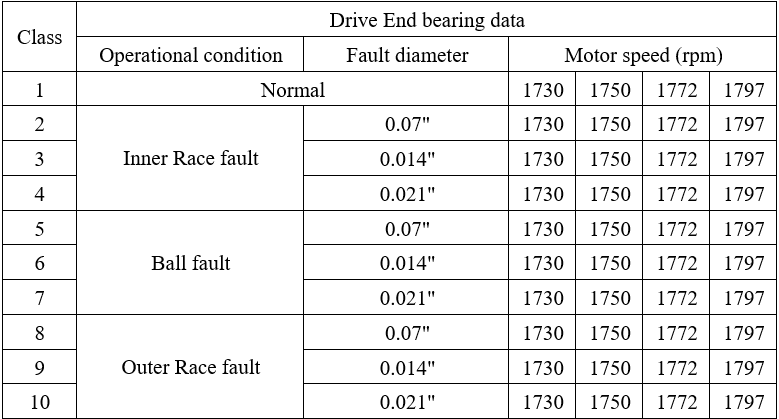}
\end{table}

\section{Performance Evaluation}

The Case Western Reserve University (CWRU) bearing test data \cite{CWRU} is used to evaluate performance of the framework \footnote[1]{The model is implemented using Python programming language and the code is available at: \url{https://github.com/Western-OC2-Lab/Similarity-Based-Predictive-Maintenance-Framework-for-Rotating-Machinery}}. The CWRU bearing test data consists of vibration signals of normal and faulty bearings. Faults ranging from 0.007 inches in diameter to 0.040 inches in diameter were introduced separately at the inner raceway, the ball, and outer raceway. Faulted bearings were reinstalled into the test motor and vibration data was recorded for motor loads of 0 to 3 horsepower (motor speeds of 1797 to 1720 RPM), and digital data was collected at 12,000 samples per second. Table \ref{dataset_table} shows operational condition, fault diameter, and motor speed of these vibration signals. According to the operational condition and fault diameter, abnormal operational conditions are classified into 9 classes as shown in the table. Thus, the dataset consists of 10 operational conditions or classes, one normal operational class and nine faulty operational classes.\

Before processing the signals, each vibration signal is divided into samples of 2000 data points each. With a sampling rate of 12000 KHz, this gives a sampled vibration signal of 0.166 seconds interval \cite{Attoui2}, which would be short enough to serve the purpose of condition monitoring and reduce computational requirements. The resultant sampled dataset consists of 3019 vibration signals. For each class, one reference signal is selected for each motor speed. This yields a total of 40 reference signals with 4 reference signals for each class, and a total of 2979 test signals.\

The next step involves corrupting reference and test signals with AWGN according to the desired SNR. Consequently, signals are denoised using WPD. \

\begin{table*}[t!]
  \caption{Accuracy results of time, frequency , and time-frequency features}
  \label{results}
  \includegraphics[width=0.9\textwidth]{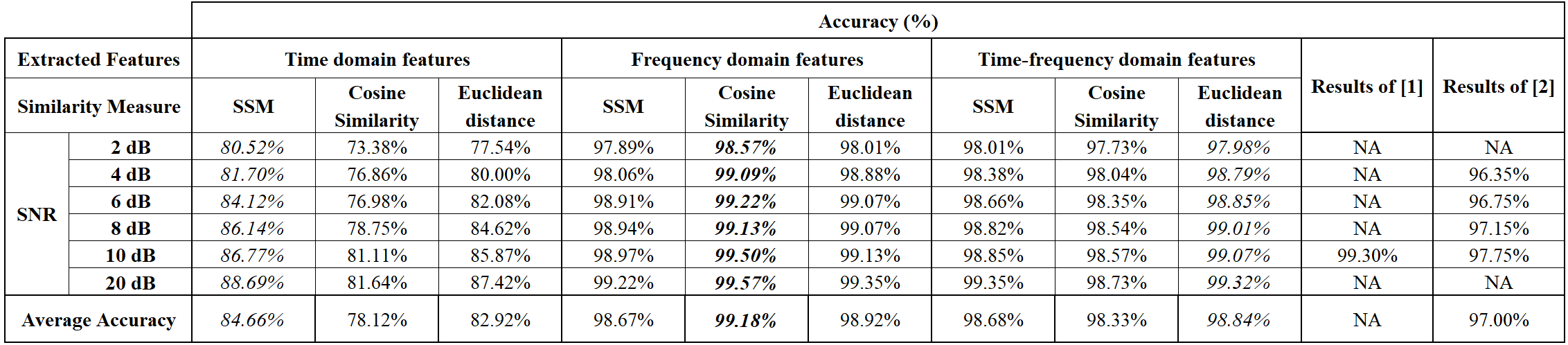}
  \centering
\end{table*}

\begin{figure*}[t!]
\centerline{\includegraphics[width=0.9\textwidth]{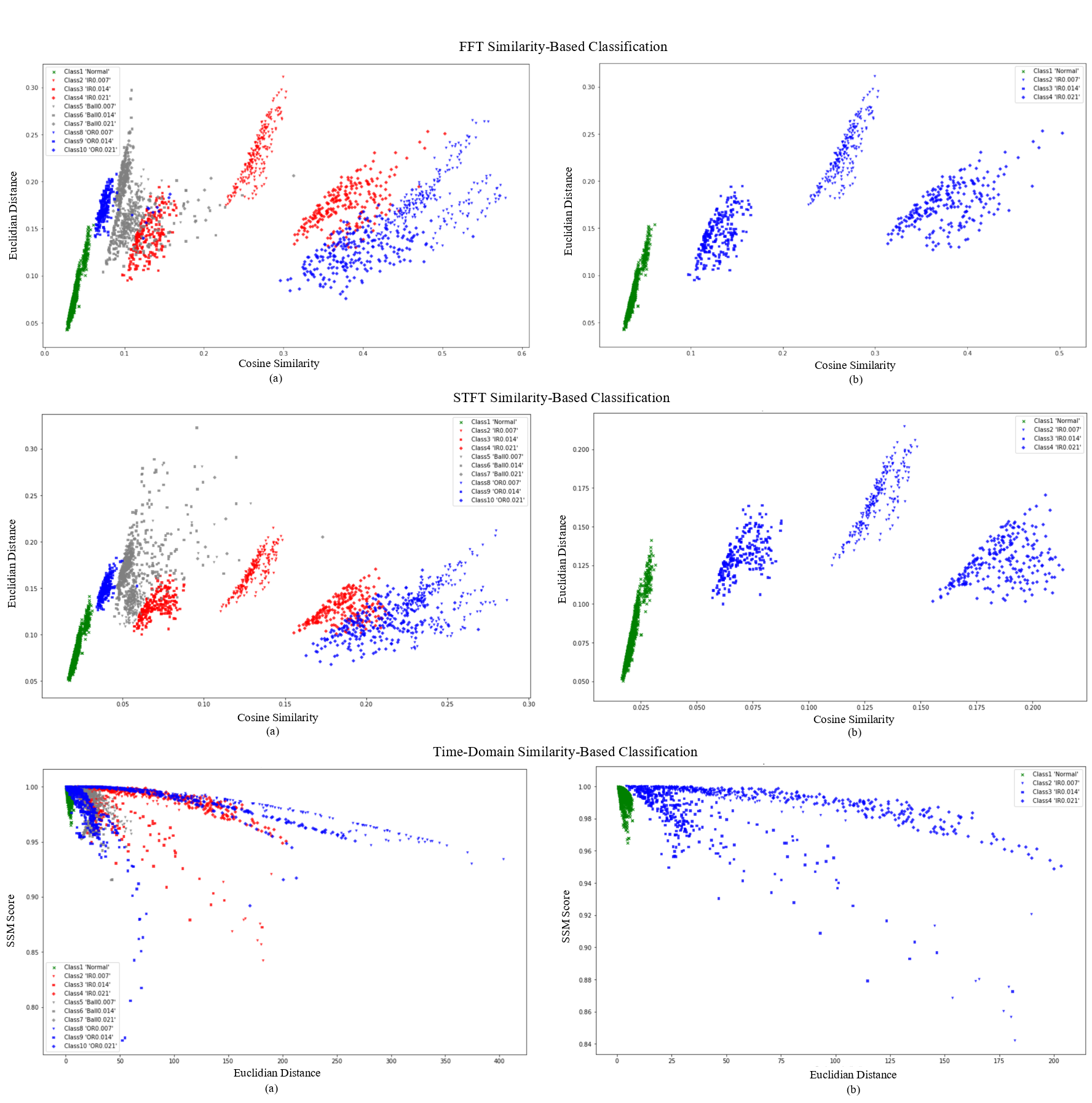}}
\caption{scatter plot shows (a) all classes and (b) 
normal and Inner Race (IR) faulty classes.}
\centering
\label{fig5}
\end{figure*}

After the denoising operation, reference and test signals are fed into the feature extraction stage. Time domain features are extracted as proposed in \cite{Nayana}. Frequency domain features, and time-frequency domain features are extracted according to settings in \cite{Attoui1} and \cite{Attoui2}, respectively. After the feature extraction stage, reference signals of the same class are combined into one reference signal by taking their mean. This will reduce computational requirements during next stages, Also, it is found that averaging same-class reference signals at this stage improves the accuracy compared to averaging them before the feature extraction stage. This improvement in accuracy can be justified as follows: Averaging same-class reference signals into one signal after the feature extraction stage will maintain feature-related characteristics of each averaged signal, and will reduce the probability of error in the similarity-based classification.

\section{Results and Discussion}
The performance of the framework is evaluated in terms of classification accuracy which quantifies the correctly classified test samples as a percentage of total test samples. Table \ref{results} shows accuracy performance under different SNR conditions. As shown, the classification accuracy of the three similarity measures with time-domain, FFT, and STFT features is evaluated at SNR levels of 2 dB, 4 dB, 6 dB, 8 dB, 10 dB, and 20 dB. The aim here is to evaluate the effectiveness of  the extracted features and the employed similarity measures under extreme, moderate, and low noisy conditions. The first observation in the results is that STFT and  FFT features have a similar performance; both achieved very high accuracy ($>98\%$) under all noisy conditions. It is noticeable that even though FFT has no time resolution compared to STFT, it achieved slightly better performance. From Fig.\ref{fig3}.a, it is clear that STFT spectrogram does not reveal any  obvious transitions in the frequency within the captured signal's length of 0.166 seconds. Thus, the time resolution aspect of STFT didn't add more discriminative characteristics to the STFT features compared to the FFT features. Instead, it may caused slight redundancy in the extracted features.
The second observation is that time-domain features have lower performance compared to STFT and FFT. In contrast to frequency contents, the statistical time properties extracted from vibration signal's waveform have low sensitivity to the variations in operational conditions; especially with small lengths of captured signal \cite{Nayana}. Fig.\ref{fig5} demonstrates the effectiveness of FFT and STFT features in similarity-based classification for vibration-based condition monitoring. The figure shows two-dimensional scatter plots of test samples with time-domain, FFT, and STFT features. For each feature type, the samples are represented on the $X-Y$ plane by their scores of the two best performing similarity measures. Specifically, for FFT and STFT features, cosine similarity along with Euclidean distance are used to plot the test samples on the $X-Y$ plane. While for time-domain features, SSM score and Euclidean distance are used. As shown, unlike, time-domain features, FFT and STFT features have large difference margins in terms of similarity scores between the operational classes.This demonstrates the high discriminative capabilities of FFT and STFT features between the different operational classes compared to time-domain features.\
Regarding the effectiveness of the applied similarity measures, all three measures are effective in classifying different operational conditions with FFT and STFT features. While the three similarity measure have a very similar performance, the cosine similarity with FFT features achieved the best performance compared to SSM and Euclidean distance with an average accuracy of $99.57\%$. Table \ref{results} also compares the obtained results with recent related work \cite{Attoui1}\cite{Attoui2}. As shown in the table, the obtained FFT results are in conformance with results of \cite{Attoui1} where FFT is used to extract the features. However, the obtained STFT results are better than results of \cite{Attoui2} even though STFT with the same settings is used in this paper. In contrast to signal denoising approach that is used in in this paper, the signals in \cite{Attoui2} are decomposed using WPD and reconstructed using  only  characteristic  fault  frequencies  and most impulsive frequency bands (MIFBs) only. As a result, some of the useful frequency components could be filtered out; which would explain the performance gap between the two approaches.

\section{Conclusion}
In this paper, a data-driven PdM framework using similarity-based classification is introduced for condition monitoring of rotating machinery. The framework addresses real-world situations of limited availability of labeled data by applying a similarity-based classification where one labeled sample for each operational condition is enough to carry out the classification task. The performance was investigated under different noisy conditions, and WPD was employed to denoise the signals and increase system robustness against the noise. Nine settings of feature type-similarity measure combinations were evaluated under different SNR levels. Experimental results corroborate the effectiveness of similarity-based approaches in vibration-based condition monitoring. Further, results demonstrate the capability of similarity-based approaches in achieving very high accuracy with moderate computational requirements compared to machine learning-based methods. Moreover, the results indicate that using FFT features with cosine similarity would lead to better performance compared to the other settings.

\section*{Acknowledgment}

This work was funded in part by National Research Council Canada under Project no.: AM-105-1.

\ifCLASSOPTIONcaptionsoff
  \newpage
\fi

\end{document}